# Conceptual Modeling Applied to Data Semantics

Sabah Al-Fedaghi[*]

*Computer Engineering Department*
*Kuwait University*
*Kuwait*

salfedaghi@yahoo.com, sabah.alfedaghi@ku.edu.kw

*Abstract* - **In software system design, one of the purposes of diagrammatic modeling is to explain something (e.g., data tables) to others. Very often, syntax of diagrams is specified while the intended meaning of diagrammatic constructs remains intuitive and approximate. Conceptual modeling has been developed to capture concepts and their interactions with each other in the intended domain and to represent structural and behavioral features of the modeled system. Conceptual modelers must ensure that a model properly embodies semantics in reality. This paper is a venture into diagrammatic approaches to the semantics of modeling notations, with a focus on data and graph semantics. The first decade of the new millennium has seen several new world-changing businesses spring to life (e.g., Google and Twitter), that have put connected data at the center of their trade. Harnessing such data requires significant effort and expertise, and it quickly becomes prohibitively expensive. One solution involves building graph-based data models, which is a challenging problem. In many applications, the utilized software is managing not just objects as well as isolated and discrete data items but also the connections between them. Data semantics is a key ingredient to construct a model that explicitly describes the relationships between data objects. In this paper, we claim that current ad hoc graphs that attempt to provide semantics to data structures (e.g., relational tables and tabular SQL) are problematic. These graphs mix static abstract concepts with dynamic specification of objects (particulars). Such a claim is supported by analysis that applies the thinging machine (TM) model to provide diagrammatic representations of data (e.g., Neo4J graphs). The study's results show that to take advantage of graph algorithms and simultaneously achieve appropriate data semantics, the data graphs should be developed as simplified forms of TMs.**

## I. INTRODUCTION

> We were blown away by the idea that it might be possible to replace the tabular SQL semantic with a *graph*-centric *model* that would be much easier for developers to work with when navigating connected *data*. [1] (Italics added)

It is a quite recent movement among philosophers, logicians, cognitive scientists, and computer scientists to focus on various types of representation systems, and much research has been focused on graph and diagrammatic representation systems in particular [2]. This paper is a venture into

---

diagrammatic approaches to the semantics of modeling notations, with a focus on data semantics. Data semantics includes representation, expressiveness, and control of data. The research task involves three main notions: data relations, graphs, and modeling, as expounded in the following three subsections.

*A. Data Relations*

Relations are omnipresent in data and in how we interact with visualization tools. Comparing, evaluating, and interpreting related pieces of information are fundamental tasks in visual data analysis but also in information-intensive work in general [3]. According to Robinson et al. [1], the first decade of the new millennium has seen several new world-changing businesses spring to life (e.g., Google and Twitter), that have put connected data—graphs—at the center of their business. Data volume will definitely increase in the future, but what's going to increase at an even faster clip is the *relationships* between individual data items [4]. Connected data is data whose interpretation and value require first an understanding of the ways in which its constituent elements are related [4].

In enterprise-content management applications, software manages not just many individual, isolated, and discrete data items but also the *relations* between them. Although we could easily fit the discrete data into such a structure as relational tables, the connected data was more challenging to store and took an extremely long time to query [1]. According to Allen [5], the simplest way to think about relationships is to write declarative sentences and isolate the nouns and verbs. All nouns are nodes, and all verbs are relationships. Such data imply a *graph*.

*B. Graphs*

Graph theory has been around for nearly 300 years and is well known for its wide applicability across a number of diverse mathematical problems and databases [1]. When Euler invented the first graph, he was trying to solve a specific problem of the citizens of Königsberg with a specific model and a specific algorithm. It turns out quite a few problems can be addressed using the graph metaphor of objects and pairwise relations between them [6].

The concept of a graph inherently concerns relationships [7] that can uncover insights about a relationship's strength and direction [8]. Here, the term "graph" refers to "graph analytics, built on the mathematics of graph theory, [which] is used to model pairwise relationships between people, objects, or nodes in a network" [8]. Graphs are a powerful tool to model relationships between data items in ways that enhance efficient retrieval as well as search queries that are specific and enable the use of more complex queries. According to Robinson et al. [1], graphs are extremely useful for understanding in fields such as science, government, and business. The real world is rich and interrelated: uniform and rule-bound in parts, exceptional and irregular in others. Gartner predicts graphs will be one of the top four enabling technologies for data and analytics. In this context, data *modeling* is the process describing an arbitrary domain as a connected graph of nodes and relationships with properties. In a database system, a data *model* shows how the system views and processes its data. For example, the relational data model presents data in terms of tables (relations).

Graphs are powerful tools for constructing *models*, and graphical representation is commonly utilized to communicate a system's functional and data flow characteristics and requirements. Graphs are used to represent conceptualizations; however, typically, dynamic features are conceptualized in a way that fails to integrate structure and dynamic features.

*C. Modeling*

People usually use models to help them understand, organize, study, and solve problems [9]. In this article, we discuss the application of graphs in modeling, specifically conceptual modeling. In modeling, the meanings of "graph" and "diagram" have only small differences; therefore, we will use them interchangeably. For example, according to IBM [10], the Unified Modeling Language (UML) *diagram* provides a visual representation of a system's aspects. Such diagrams illustrate the quantifiable aspects of a system that can be described visually, such as relationships, behavior, structure, and functionality. On the other hand, according to Neo4j basics [11], a *graph* is a set of objects and connections between pairs of the objects. The objects are known as **nodes**, and the connections are known as relationships. With these two elements, we can *solve real-world problems*.

Modeling is as old as mankind, allowing humans to represent reality to communicate, understand how something works, and devise and build more complex artifacts and systems [12]. Modeling manifests in many domains having to do with complex systems, including software engineering. We often use abstractions when attempting to model concepts, producing an abstraction of a system to be built, known as a model.

According to Booch [13], the entire history of software engineering can be seen as one of raising levels of abstraction. As such, software modeling has developed to provide concepts and mechanisms related to modeling as well as tools for editing, building, and managing models and their related artifacts. A variety of scientific and engineering fields could certainly benefit from these advances and tools [12]. In this context, according to Cabot and Vallecillo [12], broadening the horizons and aiming to realize modeling's full potential will improve modeling as a whole, opening the doors to many new and exciting opportunities for the modeling community.

A conceptual model consists of concepts used to help people know, understand, or simulate the system that the model represents [12]. These models provide representations of systems and share a set of common characteristics. Conceptual modeling, as a field of study, concerns defining formal and suitable forms of higher abstraction of the application domain to support effective and efficient development of information systems [14]. Conceptual models are considered considerably valuable in achieving effective design and implementation of advanced analytics solutions [15]. Conceptual modeling has developed to provide concepts related to modeling as well as tools for building models and their related artifacts. Conceptual modelers must ensure that their models correctly characterize the intended domain.

Multiple conceptual data models exist to support the representation of various domains and applications. For each of these conceptual models, a language is defined. Currently, the most used languages are UML and EER [16]. A recent study [17] shows that these languages have some common core entities. Additionally, UML has been used in the form of UML-oriented web services independent of any specific model. This solution has been built on top of a graph database (Neo4J) in a simple way, providing answers to the problems of representing a huge quantity of information [18].

II. DATA SEMANTICS AND DIFFICULTIES

Harnessing data sources requires significant effort and expertise, and it quickly becomes prohibitively expensive. One solution involves building *graphs* in semantics *data models*.

*A. Importance of Data Semantics*

The term "semantics" usually is related to the study of meaning. The usage of the term "semantics" has grown in popularity in areas related to information technology (e.g., the Semantic Web). Data semantics is the key ingredient in developing working solutions for data publishing, retrieval, reuse, and integration, which can be accomplished by constructing a semantics model that explicitly describes the relationships between data items in addition to their semantic types [19]. Modeling the semantics of data sources requires significant effort and expertise, and building semantic models is a challenging problem [19].

For example, in relational databases, adding relationships by embedding a table's identifier inside the field belonging to another table (foreign keys) requires joining tables, which quickly becomes prohibitively expensive. The logical data structure of a DBMS (e.g., relational) cannot satisfy data semantics because it is biased toward the implementation strategy the DBMS employs. According to Buxton [20], "Once you've got your tables set up, you can certainly store that information in tables and query it in SQL, but your SQL queries wind up being massive numbers of joins and relational databases are just not made for that kind of storage or query of query, so they tend to do very poorly." Developing data semantics helps pull all information together and draw relationships between the data pieces.

*B. Sample Approach to Data Semantics: Graph Database*

Data sources, e. g., relational databases, rarely provide a semantic model to describe their contents. Traditionally, data semantics impart meaning to data by making explicit what the data represents, i.e. the concepts (the things being modeled) and the relationships between these concepts [21]. One approach to data semantics is adopting a graph approach that is claimed to transform organizations in the "data age" [22]. Accordingly, it is claimed that relational technology models have reached their limit in view of the big-data movement, which deals with a rapidly increasing volume of data sources, accompanied by soaring expectations for new insights and data-derived value [22]. The semantic graph database is a graph database (e.g., NoSQL "Not only SQL") that stores data as a network graph and use graph data model that provides a rich contextualization of relationships to inform data analytics. Query engines in semantic graph environments have query speed and the sheer amounts of data and techniques that maximize the discovery of relationships across a unified semantic model, enabling the parsing of billions of semantic statements each second. The use and analytic insight of these databases' graph-aware approach translates into higher rates of productivity, allowing organizations to optimize their manpower [22].

*C. Problem Discussed in This Paper*

Complex and dynamic relationships in connected data require the ability to understand and analyze graphs [1]. Structural correspondence between a diagrammatic depiction and semantic content plays "a crucial role in both interpretation and inference processes with the representations" [23]. According to Morgan [24], quoting Gartner, "We needed a truly elastic computing environment in order for graph [analytics] to really work. And we needed an elastic compute environment to figure out what meaningful boundaries were needed in a graph." There is a great need to understand and appreciate the nature of graphs/diagrams.

In this paper, we claim that *current graph/diagrammatic representations do not completely depict their underlying semantics or provide a basic static structure with elementary dynamic features, creating a conceptual gap that sometimes causes misinterpretation*. Additionally, current diagrammatic representations (e.g., graph databases) mix *static* description with *dynamic* specifications. In this paper, we introduce this issue in terms of diagrammatic representation of data (e.g., tabular SQL) using several examples, emphasizing graph databases and tabular SQL semantics.

Although graphs/diagrams are important tools to make data processing more effective in many models, of which we will provide examples, the handling of data semantics is inadequate. Specifically, the deficiency discussed in this paper can be summarized as follows.

*Concerning mixing static-level and dynamic-level modeling in which objects (particulars) are present in the abstract level. The issue is how to represent an object diagrammatically and the difference between objects and events, usually discussed based on the ontological distinction between endurants and predurants*

To expose such a problem, we adopt a methodology for expressing diagrammatically the "meaning" in software and a conceptual data model called a thinging machine (TM), briefly described in the next section. Section 4 contains an elaborate example that demonstrates various features of the TM model. In the paper's remaining sections, we discuss the problem mentioned above by analyzing examples from the published research literature.

III. A GLIMPSE OF THE THINGING MACHINE MODEL

This section contains a glimpse of basic TM notions taken from more elaborate materials in a very recent paper [25].

A TM model is based on thimacs (things/machines). The thimac has a dual mode of being: the thing side and machine side. In TM modeling, thingness and machinery cannot be separated, but it is often convenient to focus on one or the other aspect. The TM machine has five actions: create, process, release, transfer, and receive (see Fig. 1). For simplification, we assume that all arriving things are accepted; thus, we can combine arrive and accept stages into the receive stage. Fig. 2 shows a thimac in terms of a thing and a machine.

Additionally, the TM model includes the triggering mechanism (denoted by a *dashed arrow* in this article's figures), which initiates a (non-sequential) flow from one machine to another. Triggering is a transformation from flow of one thing to flow of a different thing (e.g., flow of electricity triggers flow of hot water).

The TM model is based on a fundamental distinction between static and dynamic models, and within the dynamic model, it differentiates between instances and events.

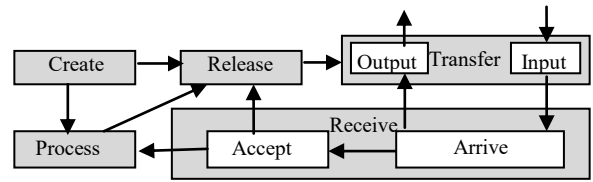

Fig. 1. Thinging machine.

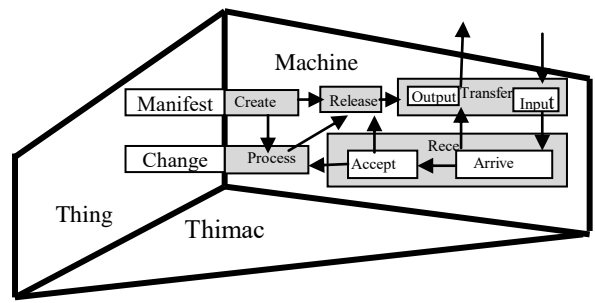

Fig. 2. Illustration of the thimac as a thing and machine.

## IV. TM Modeling Example

As a sample of the state of the art in modeling business-process analysis [26], in a 2022 study, Missikoff [27] presents a methodology that provides "a formal grounding that guarantees the quality of the produced specifications." According to Missikoff [27], software projects are among the most difficult engineering undertakings, difficult to be managed and carried out to produce a final product with the required characteristics within the time and budget originally planned. Although the significant advances in software engineering specifically requirement engineering, software projects still face a number of problems, such as the problem of misalignment between business needs and the services the information system offers. One of the main causes of problems in business process analysis resides in the collection of business and user requirements, which often yield poor requirement specifications and modeling.

Business-process analysis comprises gathering and organizing domain knowledge for later use in software development. The quality of the developed information system largely depends on how such an analysis has been carried out and the quality of the produced requirement-specification documents [27].

Then Missikoff [27] proposes a knowledge framework organized in eight knowledge sections aimed at helping the business expert conduct the analysis, eventually yielding a business-process analysis ontology. We demonstrate this framework with an example of a home delivery pizza business. It involves interviews with pizza shop operators (e.g., "*Mary connects to the PizzaPazza Web site and places his order of two Napoli pizzas, also providing payment. On the arrival of Mary's order at PizzaPazza, John, the cook, puts the order on the worklist...*") [27]. Fig. 3 shows an excerpt of the functional Pizza Shop class diagram and the ontology such a framework produces.

Such a real-life process of requirement specifications and modeling provides an opportunity to demonstrate various features of the TM model. In a TM, the process's initial phases are based on flows of things in machines. *Flow* refers to a thing's "entrance" into various machines; for example, a physical artifact's movement along an assembly line eventually leads to its *arrival* in two robots, as Fig. 4 shows. In such an approach, a designer forms a train of a flow that follows another, connected by flows (including triggering). Accordingly, guidelines for devising a TM design for a particular problem are as follows:

− "Pick up" a thing
− Follow its flow through machines until
− It triggers another thing, which leads to a new stream of flow to follow, etc.

For example, in the pizza example, the flow of an *order* leads to adding the order in the *work list*. From the work list, the flow of an order leads to flows of *dough* and *ingredients* to prepare a *raw pizza* that flows to the *oven* and then to the *delivery person*.

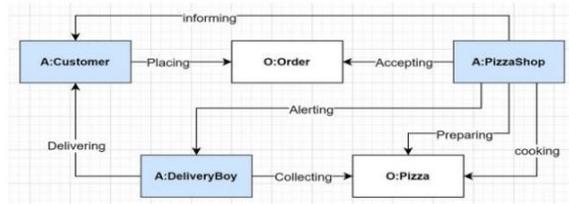

Fig. 3 Excerpt of functional PizzaShop-class diagram and ontology (from [27]).

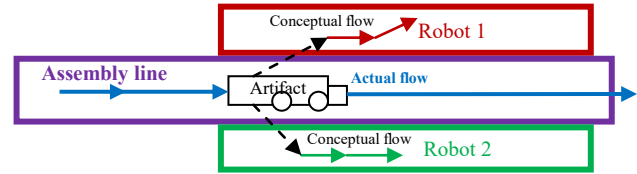

Fig. 4. Illustration of two conceptual flows triggered by an artifact's actual flow.

### A. Static model

Fig. 5 shows the TM description (static model) of the first phase of the pizza shop. In Fig. 5, the customer creates (the pink number 1) an order, which corresponds to a class in UML terminology. As the figure shows, the system receives the order (2) to trigger sending an invoice (3), which the customer processes (4) to send a payment. Upon the payment's arrival (5), the pizza system sends it to the bank (7), which sends a response to the system (8). If the payment is *not OK*, a rejection message is sent to the customer (9); otherwise, the order is sent to the work list module (10). In the work list module (11), the new order is added to the list (12) (e.g., the FIFO queue).

Simultaneously, another process begins with the work list (12). The work list is processed, and if it is empty, it waits (13), and if it is not empty, the list is processed (14) to extract one order to be processed (15). Note that the result of processing the list triggers transfer/receive because the order already exists in the list, analogous to receiving a passenger when an airplane lands and is processed to let the passengers out.

Now the order is ready to be implemented by preparing (16) the dough and ingredients. However, such a preparation process would start only if there is a place for the new pizza in the oven (17). Assuming that the oven is not full, the preparation process starts (16). The dough (18) and ingredients (19) are combined (20) to produce a raw pizza (21). The raw pizza is put in the oven, which has a counter (22) counting the current number of pizzas in the oven. Processing the raw pizza in the oven creates a cooked pizza (23), which is moved to the delivery person (24). The delivery person also has a copy of the order. Then the delivery person goes to the customer (26 and 27) and delivers the pizza (28).

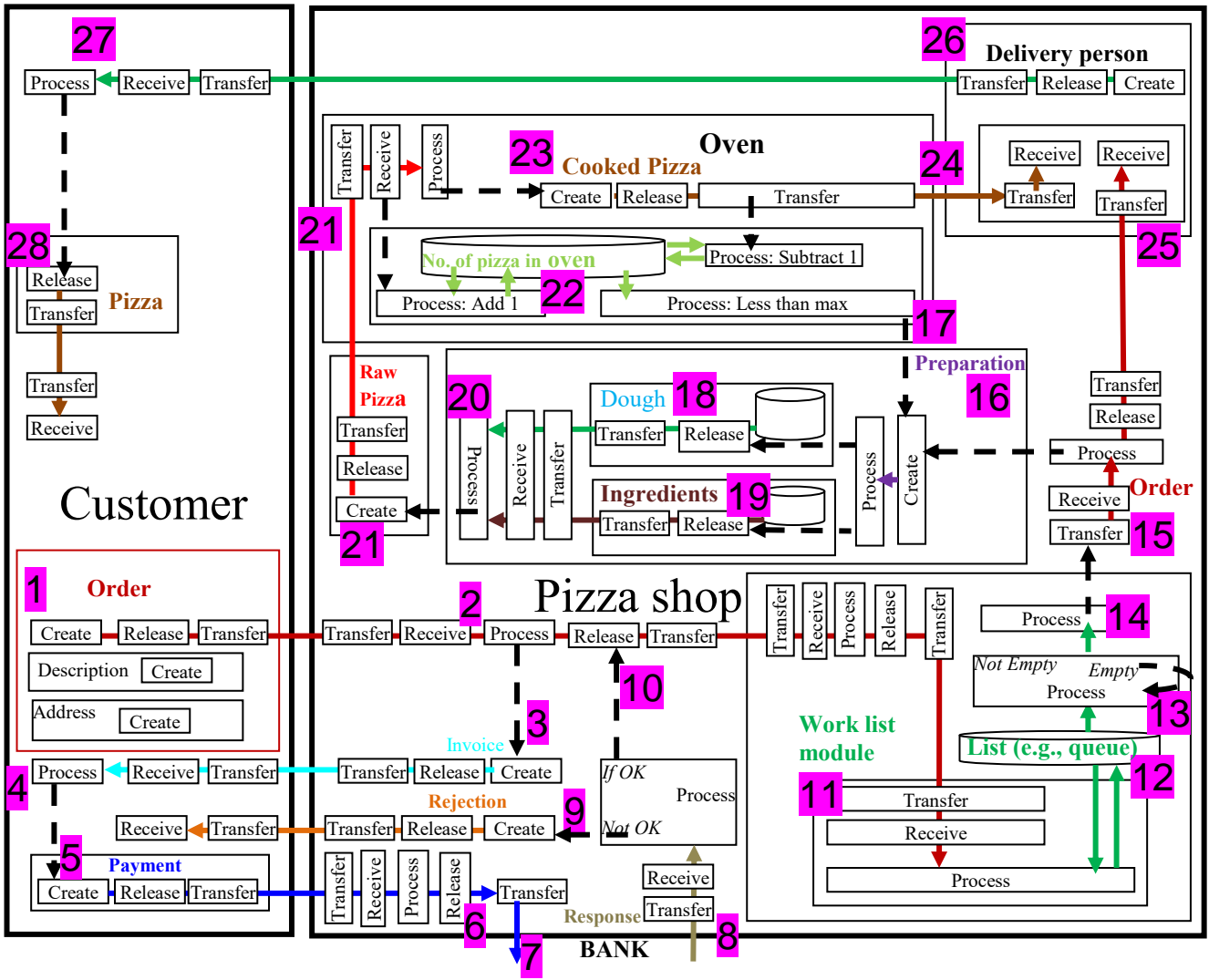

Fig. 5 Static model.

*B. Dynamic Model*

To develop the pizza system's behavior, we have to identify events. An event in TM is constructed from a subdiagram of the static model (called a region of the event) injected with time.

For example, Fig. 6 shows the event *Delivery person takes the pizza to a customer*. For simplicity, we will represent events by their regions. Accordingly, we specify the following events over the static description as follows (See Fig. 7).

Event 1 ($E_1$): An order is created.
Event 2 ($E_2$): The order is sent to the pizza shop.
Event 3 ($E_3$): An invoice is sent to the customer.
Event 4 ($E_4$): The customer sends a payment, which goes to the bank for verification.
Event 5 ($E_5$): The bank response is processed.
Event 6 ($E_6$): The payment is not OK; therefore, a rejection message is sent to the customer.
Event 7 ($E_7$): The payment is OK; therefore, the order is sent to the work list module.
Event 8 ($E_8$): The order is added to the work list.

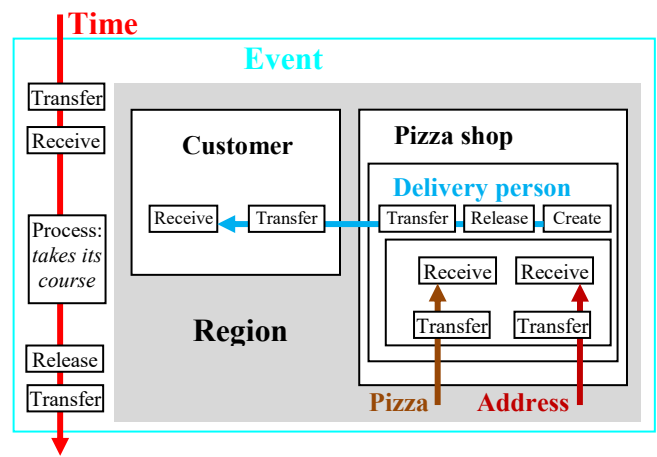

Fig. 6 The event *Delivery person takes the pizza to a customer*.

Event 9 ($E_9$): The work list is processed.
Event 10 ($E_{10}$): The work list is empty, so wait for an order.
Event 11 ($E_{11}$): An order is fetched from the work list.

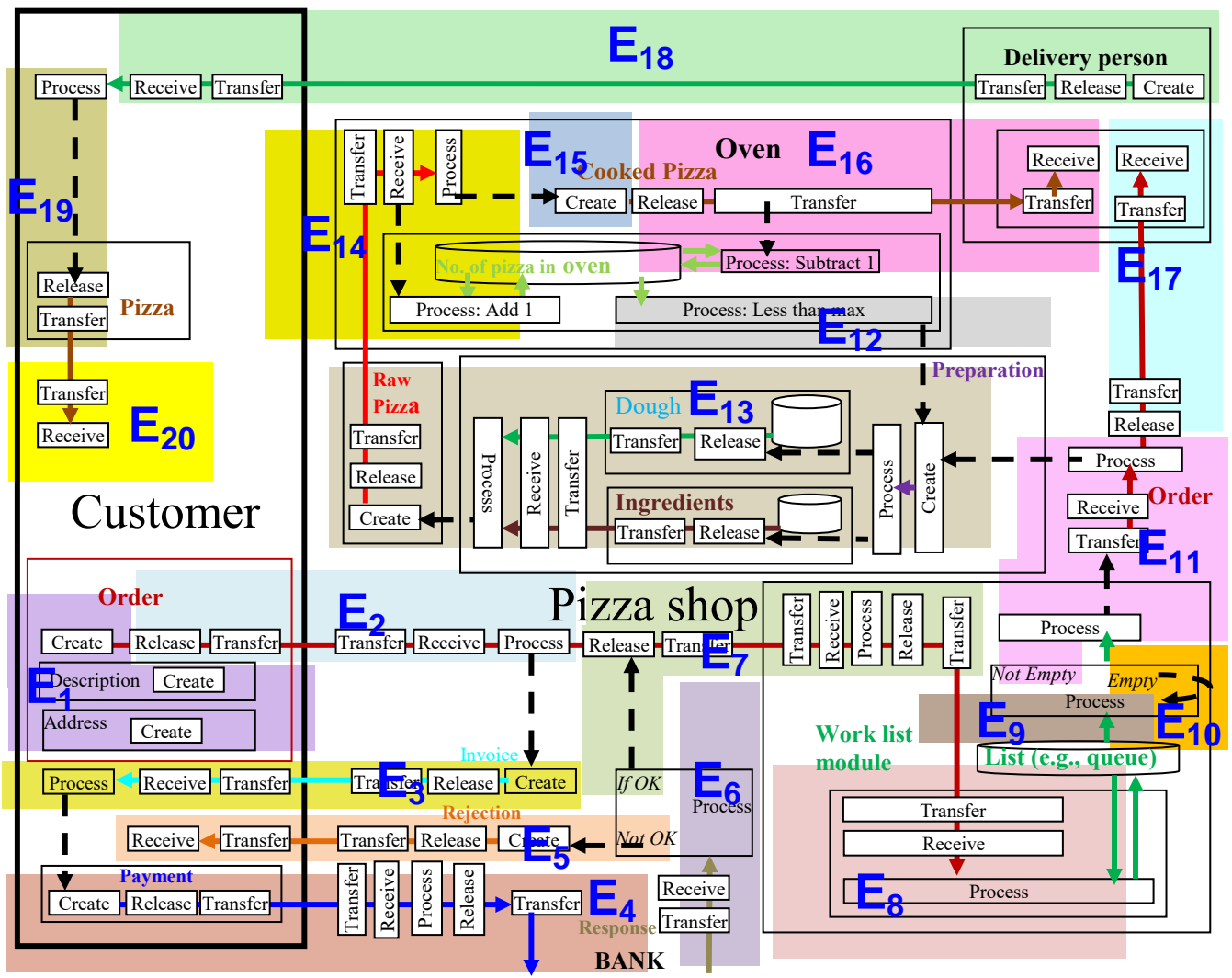

Fig. 7 Dynamic model.

Event 12 ($E_{12}$): The oven is ready to receive a pizza.
Event 13 ($E_{13}$): The preparation procedure is activated to produce a raw pizza.
Event 14 ($E_{14}$): The raw pizza is put in the oven, and the number of pizzas in the oven is updated.
Event 15 ($E_{15}$): The pizza is cooked.
Event 16 ($E_{16}$): The pizza moves from the oven to the delivery person, and the number of pizzas in the oven is updated.
Event 17 ($E_{17}$): The order is sent to the delivery person.
Event 18 ($E_{18}$): The delivery person takes the pizza to the customer.

Event 19 ($E_{19}$): The delivery person takes the pizza to the customer.
Event 20 ($E_{20}$): The customer receives the pizza.

Figs. 8 and 9 shows the pizza system's behavior in terms of two independent processes that run in parallel:

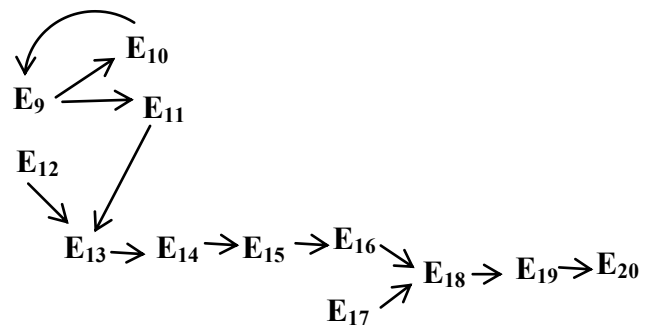

Fig. 9 Implementing a pizza order.

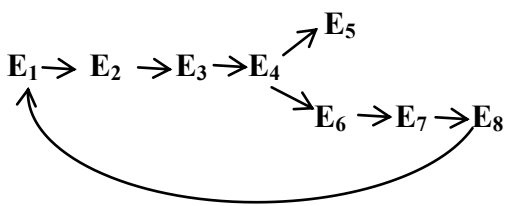

Fig. 8 Adding a new order to the work list.

adding an order to the work list and executing the pizza order. Many other details can be added to the TM model (e.g., more than one delivery person, generating an alarm when a pizza is ready, etc.). The TM model could be simplified if desired. For example, the static mode (Fig. 5) could be simplified by assuming that the direction of the arrows is sufficient to indicate the direction of flow, thus eliminating the actions release, transfer, and receive.

V. CASE STUDY 1: DIAGRAM-BASED DATA REPRESENTATION

According to Diskin and Wolter [28], people like drawing pictures to explain something to others or to themselves. When they do so for software system design, they call the pictures diagrammatic models. Very often, syntax of diagrams is specified while the intended meaning of diagrammatic constructs remains intuitive and approximate (and sometimes fuzzy to the level of becoming a modeling placebo). With the rapid invasion of model-centric trends in the software industry, models rather than code have become the primary artifacts of software development, with code to be generated directly from models [28]. In the case of database graphs, Sasaki et al. [4] asserts that one doesn't need to understand the mathematical graph theory to understand graph databases because they're more *intuitive* than relational databases. Each node in a graph represents an *entity*, and each relationship (edge) represents how two nodes are associated. As an example, Sasaki et al. [4] presents in Fig. 10 a small slice of Twitter users, represented in a graph data model. In Fig. 10, Billy and Harry follow each other, as do Harry and Ruth, but although Ruth follows Billy, Billy hasn't (yet) reciprocated.

We claim that such a diagram's semantics are problematic. From the TM point of view, this representation blurs the ontological framework. Gaining clarity about graph notations is a major priority in the field of data semantics.

A. *Problem: Distinguishing Endurants and Predurents.*

Fig. 10 superimposes two levels of representation in one diagram: (a) static level of classes (UML term), indicated in the figure as "User", and (b) objects indicated as "Billy," "Harry," and "Ruth." Often, identifying multiple levels of representations leads to a better understanding of the state of affairs.

According to Wagener [29], "a class is like a blueprint that describes all the properties an object will contain. We can't open the door and walk into a blueprint." On the other hand, objects correspond to things in the real world, carry the same names as those things carry in the real world, and should interact like those objects in the real world. Zeil [30] states, "I'm rather fond of the 'kick it' test. If you can kick it, it's an object." An object is a concrete representation of an abstraction, the class.

TM modeling occurs on two levels (see Fig. 11), staticity and dynamics. The static model (e.g., Fig. 5 in the pizza shop example) involves spatiality and actionality. The dynamic level includes events, instances, and behaviors (see Fig. 11). The static description represents the (conceptual) space/actionality-based description. We start static modeling by capturing *potential* activities (expressed in terms of the five TM actions) in reality. In the diagram by Sasaki et al. [4] in Fig. 10, "user" is an abstract term. The next subsection clarifies the notion of TM actionality.

B. *First TM level: Static (Potential) Actions*

Note that the notion of *action* is a fundamental TM concept. We classify the five TM actions under the term "actionality" and relate them where, at the dynamic level, the behavior resides. What we call actionality (the five generic TM actions) is not processability, a notion derived from the term "PROCESS," which is mixed with the notions of event [31] and dynamic behavior. PROCESS is said to describe events and behavior, but such a claim is not accurate because the *input-process-output* construct does not explicitly include time. In TM, actionality is static notion that embeds the potentialities of events, instances, and behavior, which appear when time is added to the static model. In a TM, a thimac in the static description exists/appears in the system as a thing and as a machine but without behavior (a time-oriented notion).

The static model gains behavior through the time-oriented notion of events and instance. An event is formed from

(a) A subthimac (subdiagram of the static model, called a region), which has specific spatiality (boundary) and a machinery of actionality, and

(b) Time.

In a TM, actions at the static level are constructs signifying the thimac's mechanical form. In the static model, the five actions do not reflect dynamism.

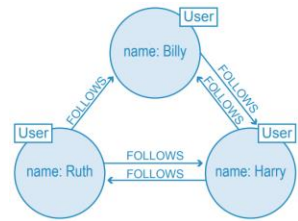

Fig. 10 The connections (relationships) between users (nodes) (from [4]).

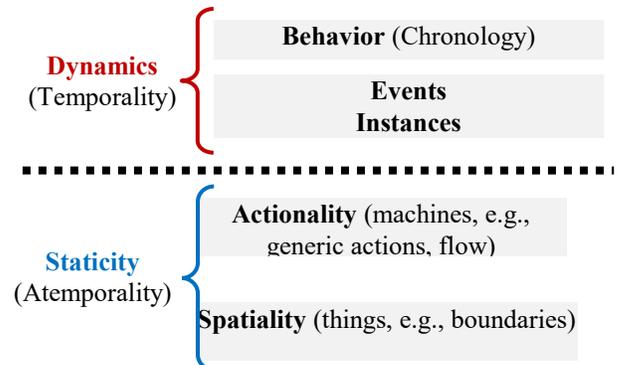

Fig. 11 TM levels.

Actionality denotes the capability (potentiality) of initiating and/or carrying out dynamism. A static thimac is the result of merging two phenomena: the spatiality of things and (potential) actionality of machines.

Continuing the example of the mechanism for *Followship* in the diagram by Sasaki et al. [4], Fig. 12 shows the TM static model or the world of *Use* (the toal thimac/diagram with three subthimacs, where user$_i$, i=1, 2, 3, is an abstract subclass or *variable*). The *Followship* is a relationship (also, a subthimac) between two subthimacs of *Use*. Two pairs of users have a mutual followship, and one pair has a directional followship. Note that the create action is not shown in some machines under the assumption that the presence of a surrounding rectangle is sufficient to indicate existence in the model. Fig. 12 can be simplified (see Fig. 13) by assuming that the direction of the arrow indicates the flow, thus eliminating release, transfer, and receive actions.

*C. Second Level: Instances and Events*

The static model shows the model's interest in the *followship* relationship in a network with three indefinite users. This is a picture of the domain of the model with three users. Applying it to reality introduces *instances* (particulars) and events. We discussed events in the pizza shop example in the previous section. Now, before presenting the dynamic model of the social network in the diagram by Sasaki et al. [4], we discuss how a TM treats *objects* (e.g., Billy) as *instances* in time.

To quote the famous John Wheeler, "Time is nature's way of keeping everything—all change that is—from happening at once." Time not only prevents all change from happening at once but also prevents all things from existing simultaneously. In TM, to specify the system's dynamics, we need to identify the portions that are susceptible to forming instances and events when injected with time. Dividing the static model causes the creation of multiple subsystems, each with its own discernable spatial region. "Dynamics" of a TM model here refers to decompositions of the static description into areas where events occur and instances (e.g., Billy, Harry, and Ruth) exist. At this point, we need to discuss the ontological separation of static and dynamic notions.

*D. Ontological Analysis of Objects (Particulars, Instances)*

Ontology research has attracted increasing attention in software engineering. Ontologies are useful tools for capturing semantics and facilitating shared understanding among fields. An ontology is an "engineering artifact" [32] representing a particular phenomenon the way we perceive reality, including changes in time. Because we are interested in the diagram relations (e.g., followship) among objects (e.g., Billy, Harry, and Ruth), in this regard, there are two main theories about the *persistence* of objects through time: endurantism and perdurantism.

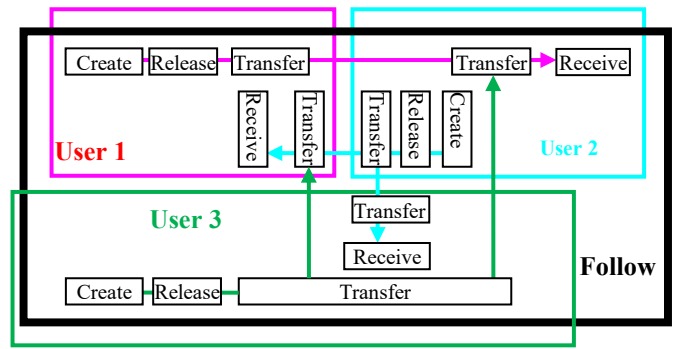

Fig. 12 The static model.

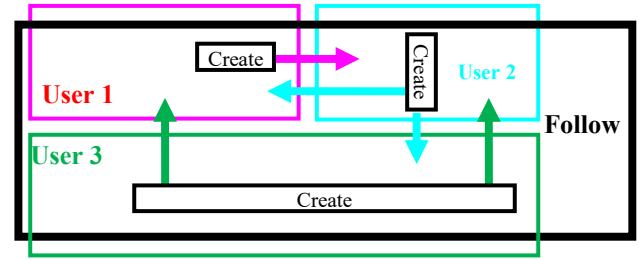

Fig. 13 Static model simplified by assuming that the direction of arrow represents the flow.

The distinction between endurants and perdurants plays a prominent role in constructing ontologies in information science. In philosophy, this distinction is conventionally introduced to discuss whether in their *persistence* through time *things* are strictly identical at every moment (endurants) or persist by unfolding different parts of themselves in time (perdurants) [33] (See Fig. 14 - table).

Endurants (also called continuants), such as objects (e.g., Aristotle's rock), have only *spatial parts* and wholly exist at each moment of their existence. Continuants are characterized as entities that are "in time," and they are "wholly" present at any time in their existence. A continuant is that which "continues to exist throughout some limited or unlimited period of time, during which its inner states or its outer connections with other continuants may be altering or may be continuing unaltered" [34]. Perdurants (also called occurrents), such as processes (landslide) and events (instance of volcano eruption), have temporal parts and exist only partly at each moment of their existence [35].

It is typically claimed (e.g., Johnson [36]) that continuants *continue* and occurrents *occur*. He claims that we cannot say that continuants occur, but this claim has been disproven [34].

| Endurant | Perdurant |
|---|---|
| 1. Objects have only spatial dimensions | 1. Objects have spatial and temporal dimensions |
| 3. Objects are viewed from the present. The default is that statements are true now. | 3. Objects from the past, present, and future all exist. |
| 4. Objects do not have temporal parts. | 4. Objects extend in time and space and have temporal and spatial parts. |

Fig. 14 Some differences between endurantist and perdurantist approaches (partial, from [37]).

Continuants can be viewed as 4-dimensional objects that are extended in time just as events are. TM modeling claims continuants are a type of occurrent, which we will call "instances."

Modeling approaches, such as entity relationship (ER) and object oriented, follow the endurantist paradigms and assume that objects have three spatial dimensions and exist in full at each moment of their lifetime [35]. Hales and Johnson [35] argued that endurantism is poorly suited to describe objects' *persistence* in a world governed by special relativity. Perdurantism, on the other hand, fits our current scientific understanding of the world [35].

In TM, endurants are static thimacs that possibly become *actual* when they become temporal. In TMs, endurants must have a region (subdiagram of the static description) (e.g., at least the action *create*), and we will use the term "instance" to distinguish them from events when time is included in their structure. Therefore, the dynamic level of a TM (i.e., events level) would result in either events or instances (See Fig. 15). Instances that continue to exist through time may be called objects (See Fig. 16).

Time is an external force that *realizes* thimacs in a static model. Realization (an object's or event's existence) is modeled in terms of time attached to a static region. Realization also indicates the object's active (e.g., creating, processing) *presence* in the model. Static thimacs are indeterminate when they are regarded without any reference to their existential correlations (events and instances) — they are ontologically representations in themselves, apart from existence, in relation to events and instances. They (static thimacs) stay forever in their unrealized and indeterminate state.

**Example** (From [34]): David Lewis (1986) argues that objects cannot possibly persist by enduring but only by perduring. If an enduring object could change from being straight to being not straight, then one and the same object would be straight and not straight, which is impossible. Fig. 17 (a) shows such an object's TM model. In the static model, the straight and not straight shapes are potentialities. In the dynamic model in Fig. 17 (b), the same object changes its shape. The chronology of events (not shown) guarantees that the same object survives the change:

*There is an object* (Event$_1$), *which has a straight shape* (Event$_2$). *The object's shape is processed* (event$_3$). *The object then is not straight* (Event$_4$).

Note that the object is an *instance* in the dynamic model because of time.

*E. The Nature of the Static Model*

The *static* TM model is an atemporal description of a *thing*'s overall life. It is a compound record of a thing. The *instance* of a TM thing is a kind of an event. This notion of an instance being an event is not a new one. Whitehead [38] gives the example of Cleopatra's Needle, an obelisk situated on the Victoria Embankment in London. For Whitehead,

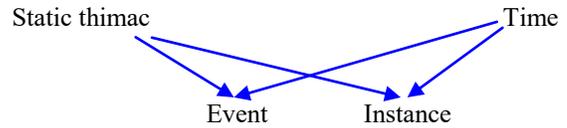

Fig. 15 Category of things at the TM dynamic level.

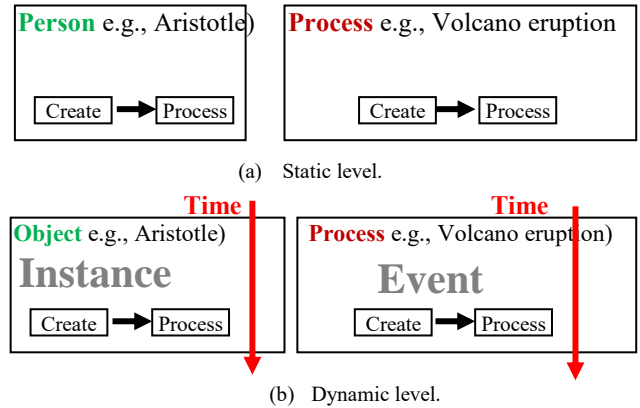

(a) Static level.

(b) Dynamic level.

Fig. 16 A TM unifies the treatment of objects and events because objects (instances) are static thimacs that embed time.

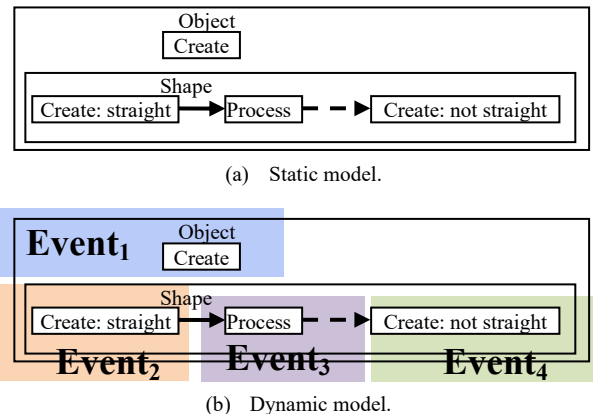

(a) Static model.

(b) Dynamic model.

Fig. 17 An object that would be straight, then not straight.

Cleopatra's Needle isn't just a solid, impassive object upon which certain grand historical events [actual changes]—being sculpted, being moved—have occasionally supervened. Rather, it is eventful at every moment. From second to second, even as it stands seemingly motionless, Cleopatra's Needle is actively happening…. At every instant, the mere standing-in place of Cleopatra's Needle is an event: a renewal, a novelty, a fresh creation. That is what Whitehead means, when he says that events—which he also calls ―actual entities‖ or ―actual occasions‖—are the ultimate components of reality. [39]

A physicist who sees such a process as a dance of electrons will tell you that daily it has lost some molecules and gained others, and even the plain man can see that it gets dirtier and is occasionally washed [38].

The behavior of such an instance (individual) as *Aristotle* is the chronology of events during his life. Accordingly, we always have to distinguish between things and actions at the static level and instances (things + time) and events (actions + time) at the dynamic level.

*F. Continuing the Followship Example*

Returning to Sasaki et al. [4], we see in the dynamic model three instances, Billy, Harry, and Ruth, and their activities, the followship relationships among them (see Fig. 18).
- There exists the instance Billy ($I_1$).
- There exists the instance Harry ($I_2$).
- There exists the instance Ruth ($I_3$).
- Billy and Harry have a mutual followship ($E_4$).
- Ruth follows Billy ($E_5$).

Here, we see that the static description provides the base for applying it to reality in terms of instances and events.

Instead of using the ad hoc graph by Sasaki et al. [4] to mark followship relationships among Billy, Harry, and Ruth, we now have an ontological apparatus (TM) for appropriate semantics. However, if there is a need for the Sasaki et al. [4] graph, it is possible to produce it from the TM dynamic model. Fig. 19 shows a first step in this direction, in which actions are eliminated from the dynamic model. Fig. 20 shows the corresponding behavior model in terms of the chronology of events.

## VI. CASE STUDY 2: GRAPH DATABASES

According Robinson et al. [1], relational databases lack relationships. To understand the cost of performing connected queries in a relational database, consider the following example from Fig. 21, which shows a simple join-table arrangement for recording friendships. When we ask the reciprocal query, "Who is friends with Bob?"

SELECT p1.Person
FROM Person p1 JOIN PersonFriend
ON PersonFriend.PersonID = p1.ID
JOIN Person p2
ON PersonFriend.FriendID = p2.ID
WHERE p2.Person = 'Bob' [1]

The answer to this query is Alice, and Zach doesn't consider Bob a friend. This reciprocal query is still easy to implement, but on the database side, it's more expensive because the database now has to consider all the rows in the *PersonFriend* table [1].

Robinson et al. [1], then, applies of graphs to an example involving a user's purchase history which "bring together several independent facets of a user's lifestyle to make accurate and profitable recommendations." Fig. 22 shows the graph of the purchase history of a user, Alice, as connected data. The graph involves linking the user to her orders and linking orders together to create a purchase history. Fig. 22 "|provides a great deal of insight into customer behavior" [1]. We can see all the orders a user has PLACED, and we can easily reason about what each order CONTAINS. A linked list structure in the graph allows us to find a user's most recent order.

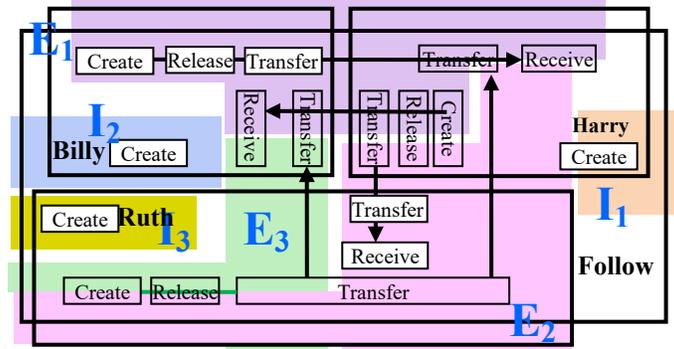

Fig. 18 The static model.

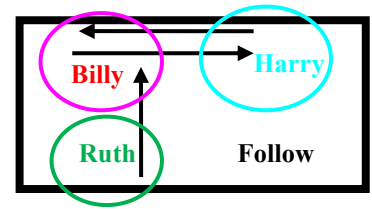

Fig. 19 Reduced dynamic model.

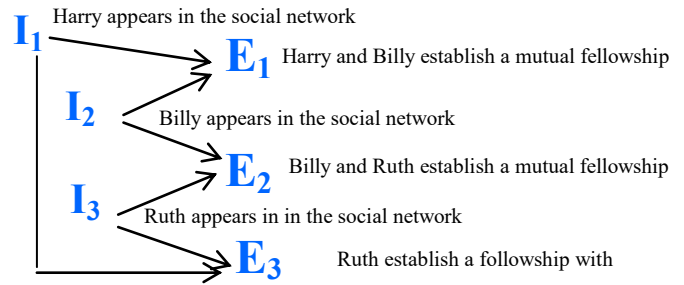

Fig. 20 The Behaviour model

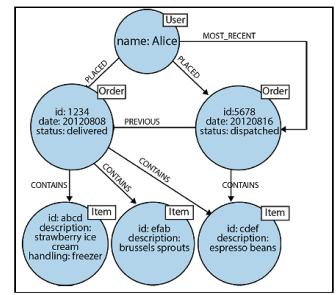

Fig. 21 Modeling friends and friends of friends in a relational database (From [1]).

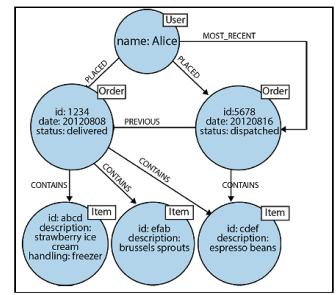

Fig. 22 Modeling a user's order history in a graph (from [1]).

As in the example in the previous section, we claim that such diagrammatic modeling is not appropriate from the semantic perspective. The diagram is an ad hoc representation that mixes levels of representations.

Fig. 23 shows the TM static model that includes the classes *customer*, *order*, and *item* and relationships between classes expressed in terms of flows. The abstract diagram is a way of understanding *patterns*, such as the user's relationship with orders and linking orders together to create a purchase history. Such an understanding requires identifying the abstract (primitive) notions of object and class to describe collections of entities and their structural relationships. Even if the interest is only in Alice, here specific orders and specific items have to be viewed in the context of abstraction: customers, orders, and items. The place of a specific customer, specific order, and specific item (instances) is not included in the static model, but their home is in the dynamic model. In the given example, *place an order* is already taken care of in the TM model as *create an order*. Additionally, *contain* (see Fig. 22) is already represented in the structure (a box inside another box) of the TM model.

Accordingly, the TM dynamic model is shown in Fig. 24 with the events/instances *a customer is created*, *an order is created*, and *an item is created*. The other events (reflecting flows) are not of interest in this example. Fig. 25 shows the behavior model in terms of creating instances. For example, Fig. 25 reflects the facts that *customer*, *order* (of customer), and *item* (of an order) have repeated instances. Therefore, if we want to represent Alice's order history diagrammatically, Fig. 26 can be built on the behavior model with full understanding of the static and dynamic levels.

## VII. CONCLUSION

According to Laux [40], the graph-based model (e.g., Neo4J, ArangoDB, JanusGraph, Amazon Neptune) has become increasingly popular, especially in the application domain of social networks, and it has been successfully applied to many other domains (e.g., medicine, drug analysis, scientific literature analysis, power, and telephone networks.). The model has been "semantically augmented with properties and labels attached to the graph elements." It is difficult to ensure data quality for the properties and the data structure because the model does not need a schema [40]. Usually, such graphs are built on the notion of objects. However, according to Laux [40], we have to deal with class things and not only with real objects.

This paper is an attempt to enhance such graph representations with an abstract schema that provides a semantic base for this type of modeling applied to data semantics. We claim that current ad hoc graphs (e.g., relational tables and tabular SQL) are problematic. Such a claim is supported by analysis that applies the TM model to sample diagrammatic representations of data. The results of study show that to take advantage of graph algorithms and simultaneously achieve appropriate data semantics, the data graphs should be developed as simplified forms of TMs.

Such a research issue needs further investigation and, specifically, application of the TM model to more examples of data graphs.

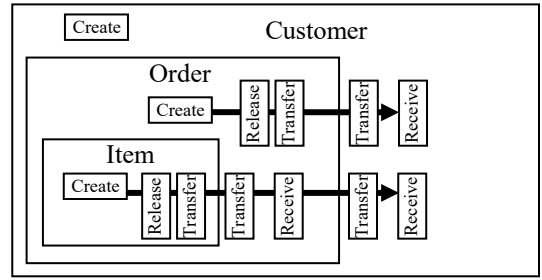

Fig. 23 Static model.

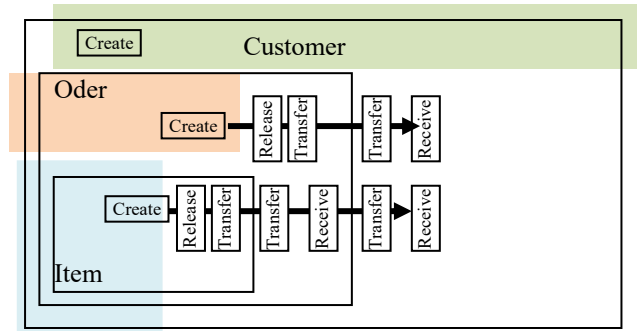

Fig. 24 Dynamic model.

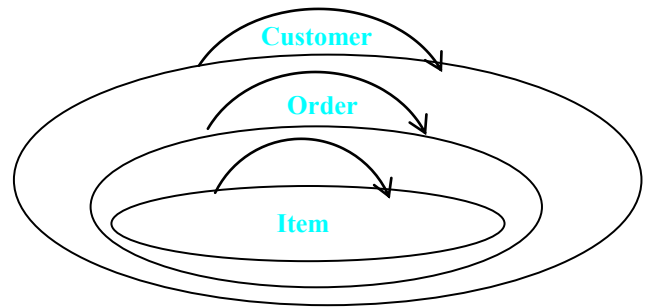

Fig. 25 Behaviour model.

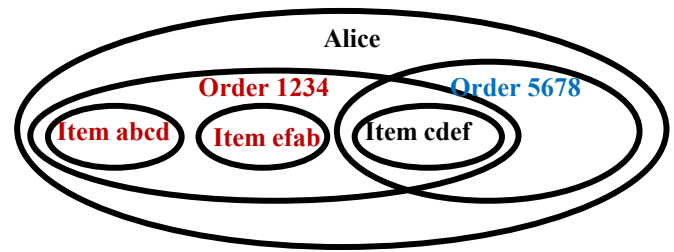

Fig. 26 Dynamic model of Alice's order history in a graph.

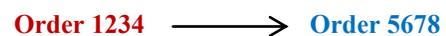

Fig. 27 Behaviour model of Alice's order history in a graph.